\documentclass[12pt]{article}
\usepackage{amsmath,amssymb}
\usepackage{amsfonts}
\usepackage[paper=letterpaper,margin=1.in]{geometry}

\newcommand{\be}{\begin{equation}}
\newcommand{\ee}{\end{equation}}
\newcommand{\bea}{\begin{eqnarray}}
\newcommand{\eea}{\end{eqnarray}}
\newcommand{\ba}{\begin{array}}
\newcommand{\ea}{\end{array}}
\newcommand{\eref}[1]{(\ref{#1})}

\newcommand{\CB}{{\cal B}}
\newcommand{\CM}{{\cal M}}
\newcommand{\CN}{{\cal N}}
\newcommand{\CO}{{\cal O}}
\newcommand{\BC}{{\mathbb C}}
\newcommand{\CP}{{\mathbb P}}

\newcommand{\Wren}{W_{\rm ren}}
\newcommand{\vev}[1]{\langle{#1}\rangle}
\newcommand{\pa}{\partial}
\newcommand{\eps}{\epsilon}

\begin{document}

\rightline{\small hep-th/0511062}
\rightline{\small DCTP-05/49}
\rightline{\small UPR-1135-T}

\vskip 0.5in

\centerline{\Large \bf Vacuum Geometry and the Search for New Physics}

\vskip 0.1in

\renewcommand{\thefootnote}{\fnsymbol{footnote}}
\centerline{\bf
James Gray${}^{1,2}$\footnote{\tt gray@iap.fr},
Yang-Hui He${}^{3,4,5}$\footnote{\tt yang-hui.he@merton.ox.ac.uk},
Vishnu Jejjala${}^{2}$\footnote{\tt vishnu.jejjala@durham.ac.uk}, and
Brent D.\ Nelson${}^{5}$\footnote{\tt bnelson@sage.hep.upenn.edu} \\
${}$ \\}

\begin{center}
{\it
${}^1$Institut d'Astrophysique de Paris and APC, Universit\'e de Paris 7, \\ 98 bis, Bd.\ Arago 75014, Paris, France \\ ${}$ \\
${}^2$Department of Mathematical Sciences, University of Durham, \\ South Road, Durham DH1 3LE, U.K. \\ ${}$ \\
${}^3$Merton College, Oxford University, \\ Oxford OX1 4JD, U.K. \\ ${}$ \\
${}^4$Mathematical Institute, Oxford University, \\ 24-29 St.\ Giles', Oxford OX1 3LB, U.K. \\ ${}$ \\
${}^5$Department of Physics \& Astronomy, University of Pennsylvania, \\ 209 South 33rd St., Philadelphia, PA 19104, U.S.A. \\ ${}$ \\
}
\end{center}

\begin{abstract}
We propose a new guiding principle for phenomenology: special geometry in the vacuum space.
New algorithmic methods which efficiently compute geometric properties of the vacuum space of $\CN=1$ supersymmetric gauge theories are described.
We illustrate the technique on subsectors of the MSSM.
The fragility of geometric structure that we find in the moduli space motivates phenomenologically realistic deformations of the superpotential, while arguing against others.
Special geometry in the vacuum may therefore signal the presence of string physics underlying the low-energy effective theory.
\end{abstract}

\newpage

\setcounter{footnote}{0}
\renewcommand{\thefootnote}{\arabic{footnote}}

\noindent\section{Introduction}
String theory is a theory of physics.
As such, it is expected to make contact with observation.
While TeV scale accelerators may conceivably, in a few years' time, directly probe strings and D-branes, the congress between high-energy theory and low-energy experiment today remains only indirect.
String theory supplies a unique framework to embed the Standard Model and its supersymmetric extensions within a consistent theory of quantum gravity.
Although it serves as an organizing principle for quantum field theories and mathematics, string theory appears so far of limited utility as an intrinsic phenomenological tool.

Various top-down and bottom-up approaches have attempted to connect string theory with such manifest properties of particle physics as the presence of three light generations, the $SU(3)_C\times SU(2)_L\times U(1)_Y$ gauge symmetry, the representations of the matter fields, and the Yukawa interactions that do such things as give mass to the fermions but do not do other things such as make protons decay.
To date, no consensus exists on string compactification to low-energies.

We must look to the minimal supersymmetric Standard Model (MSSM) itself for guidance.
Already we know that there is unexplained structure in the superpotential that signals new physics.
There is a $\mu H_u H_d$ term, for example.
The value of $\mu$, which is a mass, is around the TeV scale.
If it were any higher, this would require a fine-tuning of the electroweak sector.
If it were exactly zero, it would generate a weak scale axion.
That $\mu$ is small and positive compared to the fundamental scale demands an explanation in terms of field theory.

Just as there is unexplained structure in the superpotential, there is also hidden structure in the vacuum space, or moduli space, of the MSSM.
The supersymmetric vacuum space is locally described in the language of algebraic geometry as an affine variety which is specified in terms of relations among a basis of gauge invariant operators (composite singlets built from the matter fields).
The vacuum structure arises from supersymmetry preserving conditions, the usual D- and F-flatness constraints of the theory.
Studying the algebraic geometry of this moduli space can be viewed as a search for new physics within the MSSM.
If topological invariants of the space assume special values, this should be regarded as a feature of the low-energy theory that demands explanation in terms of more foundational physics.
The enhancement or preservation of geometric structure also motivates certain classes of deformations of theories beyond the renormalizable or low-mass level and renders others less interesting.
We therefore propose the following {\em principle for phenomenology}: any special geometry in the vacuum space should be regarded as fundamental.
In particular, if special geometry is present in the vacuum space at low-mass level, only higher-dimension operators compatible with the structure are allowed in the superpotential.
This principle has the potential to be predictive:
in a given model, some higher-dimension operators consistent with gauge symmetry must nevertheless vanish in order to preserve the special structure.
Thus any effects mediated by such operators are necessarily suppressed.

We report in this Letter on the vacuum space of subsectors of the MSSM and suggest a program to study the full MSSM and related extensions.
The fragility of topological invariants means that the special nature of the geometries depends crucially on the phenomenology of our world.
There is no known explanation for these structures within quantum field theory.

\section{The MSSM: D-terms and F-terms}
To parametrize the moduli space $\CM$, we must determine the set of gauge invariant operators in the theory and impose the F-flatness constraints to obtain the relations among the $n$ generators of the set.
These relations define the vacuum space explicitly as an algebraic variety in $\BC^n$.
For every solution to the F-terms, there is as well a solution to the D-terms in the completion of the orbit of the complexified gauge group \cite{lt}.
The D-flatness constraints are then simply a gauge fixing condition which eliminates redundancy in defining the initial basis.
In other words, we fix algebraic relations among the polynomial gauge invariant operators coming from D-flatness, after imposing F-flatness conditions.
These relations then constitute the defining equations of the moduli space locally as an affine variety.
We develop a novel and algorithmic way of looking at this by considering the gauge invariant operators of the D-terms as a ring map from the space of F-terms to the vacuum space.
This procedure is essentially a Gr\"obner basis computation in polynomial rings \cite{ghjn}.
As we are faced with a gauge theory with a complicated set of F- and D-equations, computationally intensive methods are required.
The algorithm is implemented on the computer using the algebra package Macaulay 2 \cite{mac} and is applicable to finding moduli spaces of $\CN=1$ gauge theories with arbitrary superpotentials.

For the MSSM, the gauge invariant operators are constructed out of the $49$ matter fields.
(There are $18$ directions from $Q_i$, nine each from $u_i$ and $d_i$, six from $L_i$, three from $e_i$, and two each from $H_u$ and $H_d$.)
As there are twelve generators of the Standard Model gauge group, we must in the end obtain a complex variety of dimension less than $37$.
The gauge invariant combinations of the fields that define a basis of monomials for general gauge invariant operators are tabulated in Refs.\ \cite{ghjn}, \cite{gkm}.
There are $991$ gauge invariant combinations, so the basis is extremely overcomplete and constrained by relations among the operators.
The size of the basis and the complexity of redundancies renders the calculation of the variety describing the vacuum space of the full theory computationally intensive, though the methodology is algorithmically clear.
We first consider the (renormalizable) superpotential $W(\Phi)$ and impose the $49$ F-term relations:
\be
\frac{\pa W}{\pa \Phi^i} = 0.
\ee
The F-flatness conditions carry representations of the gauge group.
We then implement the dependencies among the gauge invariants after substituting the F-terms into the basis of operators.
The resulting set of equations defines the geometry of the moduli space $\CM$.

To illustrate the technique in detail, we focus in this Letter on particular subsectors of the MSSM.
Already, there is interesting geometric structure within these subsectors.

\section{The electroweak sector}
We restrict ourselves here to the electroweak sector of the theory.
This is accomplished by setting the vevs of the quarks $Q_i$, $u_i$, and $d_i$ to zero, consistent with the unbroken $SU(3)_C$ symmetry of the Standard Model.\footnote{
The resulting theory is anomalous.
Generally, when the fields are charged under an anomalous $U(1)$, the Green-Schwarz mechanism \cite{gs} renders the theory consistent: $\vev{D^\dagger D} = \xi$.
The Fayet-Iliopoulos term is a K\"ahler parameter, whereas the moduli space is sensitive to complex structure.
The algorithm implements D-terms through a (symplectic) quotient and is unaffected by the modification.}
The renormalizable superpotential is
\be
\Wren = \mu H_u^\alpha H_d^\beta \eps_{\alpha\beta} + \lambda_e^{ij} L_i^\alpha H_d^\beta e_j \eps_{\alpha\beta},
\label{eq:Wren}
\ee
where $\alpha,\beta$ are $SU(2)_L$ indices and $i,j$ are flavor indices.
In addition to the $\mu$ parameter, there are nine constant coefficients $\lambda$ that account for the mixing of generations.
We have for now ignored right-handed neutrinos.
The gauge invariant operators in the electroweak sector are:
\be
\ba{ccc}
H_u H_d = H_u^\alpha H_d^\beta \eps_{\alpha\beta}, &&
L H_u = L_i^\alpha H_u^\beta \eps_{\alpha\beta}, \\
L L e = L_i^\alpha L_j^\beta e_k \eps_{\alpha\beta}, &&
L H_d e = L_i^\alpha H_d^\beta e_j \eps_{\alpha\beta}.
\ea
\ee
There are $22$ gauge invariant operators in total.
The four D-terms corresponding to the generators of $SU(2)_L$ and $U(1)_Y$ and $13$ F-terms define the vacuum variety.

The dimension of the space is determined computationally using methods in algebraic geometry that we detail elsewhere \cite{ghjn}.
The vacuum space is five-dimensional and is an affine cone over a base manifold $\CB$ of dimension four.
As a projective variety, $\CB$ has degree six and is described as a (non-complete) intersection of six quadratics in $\CP^8$.
(We recall that $\CP^d$ is the space of one-dimensional complex vector subspaces of $\BC^{d+1}$.)
The equations for the variety themselves are not illuminating for our purposes, so we do not include them here.
Instead we consider physically motivated deformations of the superpotential \eref{eq:Wren} and demonstrate the intriguing relationship between phenomenology and algebraic geometry.

R-parity, a symmetry introduced to ensure the stability of the proton, is defined as $R = (-1)^{3(B-L)+2s}$, where $B$ and $L$ are the baryon and lepton numbers of the superfield and $s$ is the spin of each component field.
Suppose we lift the Higgs, meaning that the vevs $\vev{H_u}$ and $\vev{H_d}$ are constrained to vanish.\footnote{
In principle, the $\mu$ term in the renormalizable superpotential lifts the Higgs by itself.
However, because $\mu$ is of order the electroweak scale, the term produces a negligible contribution to the scalar potential.}
We can do this minimally by adding dimension four terms to the superpotential that respect the R-parity:
\be
W = \Wren + \tau (H_u H_d)^2 + \tau^{ij} (L H_u)_i (L H_u)_j.
\label{eq:rp}
\ee
The added terms in eq.\ \eref{eq:rp} are natural to consider since both arise in well-motivated contexts in which heavy Standard Model singlets are integrated out of the theory:
a singlet that generates the $\mu$-term as in the NMSSM for the first term or (famously) a right-handed neutrino to generate the canonical see-saw mechanism for the second term.
Eq.\ \eref{eq:rp} is the most general superpotential containing operators up to dimension four consistent with gauge symmetry and R-parity conservation.

With the extra terms, the vacuum space of the theory is three-dimensional.
More precisely, $\CM$ is an affine cone over a base surface $\CB$.
The geometry of $\CB$ is as follows.
$\CB$ is a complex submanifold of a K\"ahler manifold, and therefore is itself K\"ahler.
It is given by the (non-complete) intersection of six quadratics in $\CP^5$.
$\CB$ is a degree four surface with the Hodge diamond
\bea
\ba{ccccc}
&&h^{0,0}&& \\
&h^{0,1}&&h^{0,1}& \\
h^{0,2}&&h^{1,1}&&h^{0,2} \\
&h^{0,1}&&h^{0,1}& \\
&&h^{0,0}&& \\
\ea
&\;\;\;\;\;\;\;\;\;\;&
\ba{ccccc}
&&1&& \\
&0&&0& \\
0&&1&&0 \\
&0&&0& \\
&&1&& \\
\ea.
\eea
For comparison, on the left we show the Hodge diamond for a general K\"ahler manifold;
our expectation from field theory is that the Hodge numbers $h^{0,0}$, $h^{0,1}$, $h^{0,2}$, and $h^{1,1}$ are typically arbitrary integers.
The base $\CB$ enjoys a remarkably simple topological structure,
and there is no known explanation for this within the context of quantum field theory.
The four independent topological invariants are all either $0$ or $1$.
The number of vanishing topological invariants provides a measure for how special the variety is.
According to the classification of surfaces in $\CP^{d+1}$ \cite{hart}, there are only three possible degree four surfaces in $\CP^5$ with this Hodge structure.
These are easily distinguished.
One finds that the base $\CB$ is the {\em Veronese surface}.\footnote{
That the surface we find is well studied in the mathematics literature, while interesting, may perhaps only indicate that the possibilities for a three-dimensional variety arising in this way are constrained.}
That is to say, $\CB$ is the image of the degree two map:
\be
\ba{cccc}
\Sigma_2:& \CP^2 & \to & \CP^5 \\
& (s,t) & \mapsto & [1,s,t,s^2,st,t^2]. \\
\ea
\ee
The affine cone over the Veronese surface is simply the line bundle $\CO_{\CP^2}(-2)$.
(To specify notation, $\CO_{\CP^2}(-n)$ is a line bundle of degree $n$ over $\CP^2$.)

The special structure is fragile.
Suppose that we instead add other renormalizable but R-parity violating terms to the superpotential, such as the gauge invariants $LH_u$ or $LLe$.
While the implied violation of R-parity may be aesthetically distasteful in that the principal supersymmetric candidate for cold dark matter is lost, neither term (alone or in conjunction) can generate proton decay.
Furthermore, the dimensionless coefficients of these terms are only weakly and indirectly constrained by observation.
Both can contribute new supersymmetric contributions to $\mu \to e \gamma$, lepton branching fractions for mesons, and the anomalous magnetic moment of the muon.
But these constraints (for 100 GeV mass scales) are no more than $\CO(10^{-3})$ for the individual dimensionless couplings, so there is no {\em a priori} reason to forbid them from the point of view of phenomenology.
However, if we take
\be
W = \Wren + \rho^i (L H_u)_i,
\ee
the moduli space $\CM$ is exactly $\BC$.
Any deformation to the renormalizable superpotential involving $LLe$ reduces the moduli space to a point.
It is precisely when the superpotential consists only of the R-parity preserving terms that the moduli space exhibits algebro-geometric structure.
This is consistent with the principle stated in the introduction:
interesting deformations for physics are those which retain or sharpen features of the vacuum space in geometry.

As one more example of this, let us now consider adding right-handed neutrinos $\nu_i$ to the electroweak sector.
The neutrinos are Standard Model singlets and gauge invariant operators in their own right.
Incorporating neutrino Yukawa couplings and a $\nu_i \nu_j$ Majorana term to the renormalizable superpotential \eref{eq:Wren} by itself changes the moduli space dramatically.
It is again the familiar cone over the Veronese surface.
With the addition of the right-handed neutrino the renormalizable superpotential, without order four terms, gives a three-dimensional variety.
This structure is stable and persists without modification upon addition of higher dimension R-parity preserving terms.

\section{Comparison to string theory}
Special structure in the moduli space of low-energy gauge theories is known to descend from geometry in higher dimensions.
This is a recognizable feature of string theory.
In all string models a relation exists between the geometry of the low-energy four-dimensional gauge theory and the geometry of the compact manifold.
In certain constructions this relation is easy to trace.
In D-brane probe models, for example, Calabi-Yau geometries are moduli spaces of the quiver gauge theory on the worldvolume \cite{dm}.
The equation for the vacuum space as an algebraic variety is the {\em same} as the equation of the singularity whose transverse directions are probed by the D-brane;
the gauge theory captures motion on the resolved space.
Algorithms have been developed to study the field theories in detail \cite{lnv}.

Although it may be embedded in this context \cite{bjl}, the MSSM is not itself a quiver gauge theory.
Matter consists of more than just bifundamental fields arising from open string degrees of freedom.
Nevertheless, the geometric structure in the vacuum space may reflect geometric structure in higher dimensions.
The existence of special structure unexplained by field theory should provide signatures of higher-energy physics, whatever the ultraviolet completion turns out to be, but inheritances of this kind are particularly natural to string theory.

\section{Continuing the program}
We have advocated an 
algorithmic algebro-geometric approach to efficiently computing the algebraic variety of an $\CN=1$ supersymmetric gauge theory with an arbitrary superpotential.
This constitutes a major advance in tackling a fundamental problem in string model-building, and we have exhibited the value for phenomenology in the case of a toy example built from the electroweak sector of the MSSM.
We present more complicated examples elsewhere \cite{ghjn}.
Of course, the most important gauge theory to consider is the full three generation MSSM itself.
We intend to perform this calculation on a supercomputer, but for now only offer partial results.
The failure is merely one of scale ---
the technique, given sufficient computational resources, will succeed in determining the vacuum space of the MSSM as an algebraic variety.
The special features of this moduli space should stimulate top-down efforts to recover the MSSM from string theory.
It would be of interest to ask what relationship exists between the gauge theory moduli space of the MSSM and the geometry of string compactifications, particularly in D-brane models and in heterotic constructions that achieve a high degree of minimality \cite{maximus}.

We expect surprises and novel insights from a dedicated survey of special structure in phenomenologically appealing gauge theories.
We regard each renormalizable superpotential as one of a family of theories, defined by adding various combinations of higher-order operators consistent with the gauge invariance to the minimal superpotential.
Some members of these families will exhibit special structure.
Many will not, for special structure is by no means universal.
Indeed, we already observe that special structure only emerges when the MSSM electroweak sector is extended to include either right-handed neutrinos or higher-order operators and disappears completely when renormalizable R-parity violation is introduced.
We anticipate that similar behavior will apply to the MSSM as a whole.
It is clear that the superpotential (interactions) are critical in fixing the geometry of the moduli space.
At each mass level, we expect to test the kind and nature of the interactions against the geometry.
In this way, we can study the geometrical import of features of the gauge theory.
The vacuum space of the one-generation MSSM, for instance, is trivial.
Structure in the moduli space depends crucially on the existence of flavor.
It may be that special geometry prefers some number of generations.
This is a new bottom-up approach to model building.

$\CN=1$ vacuum manifolds often enjoy global isometries, continuous or discrete, that are not {\em a priori} evident from the superpotential.
For example, in D-brane probe scenarios, one could find such symmetries of the vacuum variety, and then rearrange or redefine fields in the superpotential to exhibit the hidden global symmetries explicitly \cite{fhk}.
The techniques we have presented may facilitate the search for manifestly symmetric forms of the Lagrangian, but these symmetries are logically distinct from the use and utility of geometry as a selection principle in its own right.

Geometry supplies a point of contact between old physics and new.
Selecting for structure in the vacuum space should be regarded as another tool for phenomenology.
This is not an unreasonable conclusion:
consider the analogous example of fine-tuning.
Certain theories suffer constraints from observation that can only be satisfied through cancellations between unrelated parameters to a high degree of precision.
Symmetries of the Lagrangian do not insist upon these cancellations.
(An example of this is the Z-boson mass constraint in the MSSM \cite{bdn}.)
When we encounter fine-tuning, we are moved to seek an explanation from a new physical input or from a symmetry of the underlying theory.
This has led to the quite useful notion of ``naturalness'' as a precept in physics.
The precept of special structure might prove equally valuable in guiding us to explanations of empirical observations about the Standard Model.

\medskip
\section*{Acknowledgments}
We thank Steve Martin and Miles Reid for helpful discussions.
JG is supported by CNRS and by PPARC.
YHH is supported in part by the FitzJames Fellowship at Merton College, Oxford and by the Department of Physics and the Math/Physics Research Group at the University of Pennsylvania under cooperative research agreement DE-FG02-95ER40893 with the U.S.\ Department of Energy and NSF Focused Research Grant DMS0139799 for ``The Geometry of Superstrings.''
VJ is supported by PPARC.
BDN is supported by the U.S.\ Department of Energy under the grant DOE-EY-76-02-3071.

\end{document}